\documentclass[prl,twocolumn,showpacs]{revtex4}

\usepackage{graphicx}
\usepackage{bm}
\begin{document}
\title{Circuit Theory for Full Counting Statistics in Multi-Terminal Circuits}
\author{Yu. V. Nazarov and D. A. Bagrets}
\address{Department of Applied Physics and Delft Institute of Microelectronics
and Submicrontechnology, \\
Delft University of Technology, Lorentzweg 1, 2628 CJ Delft, The Netherlands
}
\date{\today}
\pacs{73.23.-b, 72.70.+m, 05.40.-a, 74.40.+k}

\begin{abstract}
We propose a theory that treats the current, noise, and, generally, the
full current statistics of electron transfer in a mesoscopic system in a
unified, simple and efficient way. The theory appears to be a circuit 
theory of $2\times 2$ matrices associated with Keldysh Green functions. We
illustrate the theory by considering the big fluctuations of currents
in various three-terminal circuits.
\end{abstract}

\maketitle

%\begin{multicols}{2} 
%GENERAL INTRODUCTION: QUANTUM NOISE
The field of quantum noise in mesoscopic systems has been exploded
during the last decade, most achievements being summarized in a recent
review article. \cite{BlanterReview} Measurement of fractional charge in
Quantum Hall regime \cite{Frac_Charge}, noise measurements in
atomic-size junctions \cite{Cron} and superconductors \cite{Kozhe} are
milestones of the field and demonstrate the importance of quantum noise
as a unique tool to study electron correlations and entanglements of
different kinds.
A very important step has been made in \cite{Levitov} where an elegant
theory of {\it full counting statistics} (FCS) has been presented. This
theory encompasses not only noise, but all higher momenta of the charge
transfer.

%SPECIFIC INTRODUCTION: MULTI-TERMINAL
Starting from pioneering work of B\"{u}ttiker\cite{Buettiker}, a special 
attention has been paid to
noise and statistics of electron transfer in multi-terminal circuits.
The correlations of currents flowing to different terminals reveal 
Fermi statistics of electrons. These cross-correlations have been
recently observed. \cite{cross-corr} Although the noise correlations
for several relevant layouts have been understood \cite{BlanterReview},
the evaluation of FCS still encountered difficulties. For instance,
an attempt to build up FCS with "minimal correlation approach" 
\cite{BlanterSukhor} has lead to contradictions .\cite{BlanterSchomerus} 
This is unfortunate, since higher-order current correlations supply 
information about higher-order electron correlations and multi-particle
interference. This information is of fundamental importance
and can be hardly obtained by any other means.

%CENTRAL POINT
In this letter, we present a calculational scheme that allows for
easy evaluation of FCS in a multi-terminal mesoscopic systems. It is of a
great intellectual enjoyment that this scheme is a simple and universal
one. In fact, it is hardly more complicated than a conventional circuit
theory of electric transport and is based on slight extension of
Kirchoff rules to $2 \times 2$ matrix structures.

%STARTING METHOD: KELDYSH FOR MANY CURRENTS
We start by introducing current
operators $\hat I_i$, each being associated with the current to a certain 
terminal $i$. Extending the method of \cite{RefYuli} we introduce a 
Keldysh-type Green function defined by
\begin{equation}
\Bigl( i\frac{\partial}{\partial t} - \hat
H + \frac12 \bar\tau_3 \sum_i \chi_i(t)\hat I_i \Bigr) \otimes \check G(t,t')= 
\delta(1-1')
\label{green}
\end{equation} 
Here we follow notations of a comprehensive review \cite{Rammer},
$\chi_i$ are time-dependent parameters, $\bar \tau_3$ is a
$2 \times 2$ matrix in Keldysh space, $\hat H$ is the one-particle Hamiltonian 
that incorporates all information about the system layout, including boundaries, 
defects and all kinds of elastic scattering. We use "hat", "bar" and "check" to 
denote operators in coordinate space, matrices in Keldysh space
and operators in direct product of these spaces respectively. The Eq. 
\ref{green}
defines the Green function unambiguously provided boundary conditions
are satisfied: $\check G(t,t') \equiv \bar G(x,x';t,t')$ approaches the common 
equilibrium Keldysh Green functions
$\check G^{(0)}_i(t-t')$ provided $x,x'$ are sufficiently far in the terminal 
$i$.
These $\check G^{(0)}_i(t-t')$ incorporate information
about the state of the terminals: their voltages $V_i$ and temperatures $T_i$.

One can easily see by traditional diagrammatic methods \cite{Rammer} that the 
expansion of $\check G$ in $\chi_i(t)$ generates all possible diagrams for
higher order correlators of $\hat I_i(t)$ and thereby incorporates all the 
information about statistics of charge transfer. If we limit our attention 
to low-frequency limit of current correlations, we can keep time-independent 
$\chi_i$. In this case, the Green functions are functions of time difference 
only and the Eq. 1 separates in energy representation. It is convenient to 
introduce the following $\chi_i$-dependent action defined as a sum of closed 
diagrams:
\begin{equation}
\frac{\partial S}{\partial \chi_i} = 
-i t_0 \int \frac{d\varepsilon}{2\pi}{\rm Tr} \ \big( \bar \tau_3 \hat I_i  
\check 
G(\varepsilon) \big)
\label{action}
\end{equation}
This allows us to express the probability for $N_i$ electrons to be transferred 
to the terminal $i$ during time interval $t_0$
\begin{equation}
P(\{N_i\})= \int_{-\pi}^{\pi} \prod_i \frac{d \chi_i}{2 \pi} e^{-S(\{\chi_i\}) -
i \sum_i N_i \chi_i}.
\label{Probability}
\end{equation}
(Higher-order) derivatives of $S$ with respect to $\chi_i$ give (higher-order) 
moments of $P(\{N_i\})$.
First derivatives yield average currents to terminals, second
derivatives correspond to the noises and noise correlations.

%METHOD: SEMICLASSICAL FUNCTIONS

Using special properties of current operators, $\chi$-dependent terms in
Eq. \ref{green} can be gauged away.\cite{RefYuli,Wolfgang} The $\chi$
dependence of $\check G$ is thereby transferred to the boundary conditions:
the gauged Green function far in each terminal shall approach 
$\check G_i(\epsilon)$ defined as
\begin{equation}
\check G_i(\epsilon) = \exp (i \chi_i \bar \tau_3/2) \check G^{(0)}_i(\epsilon) 
\exp (-i \chi_i \bar \tau_3/2)
\label{boundary}
\end{equation}

In the present form, the Eq.~(\ref{green}) with relations 
(\ref{boundary},\ref{action}) solves the problem of determination of the 
FCS for any arbitrary system layout: one just has to find exact 
quantum-mechanical solution of a Green function 
problem. This is hardly constructive, and we proceed further by deriving a
simplified semiclassical approach. First, we note that even in its exact 
quantummechanical form the Eq. 1 possesses an important property. We consider 
the quantity  defined similar to standard definition of current density,
$\bar j^{\alpha}(x,\epsilon) \equiv \lim_{x \rightarrow x'}(\nabla'^{\alpha}-
\nabla^{\alpha}) \bar G(x,x';\epsilon)/m$.
By virtue of Eq. 1 this quantity conserves so that
\begin{equation}
\frac{\partial \bar j^{\alpha}(x,\epsilon)}{\partial x^{\alpha}} =0
\label{conservation}
\end{equation}
This looks like the conservation of particles at a given
energy. However, this relation contains more information since it is a 
conservation law for a $2\times2$ {\it matrix} current.
\begin{figure}[t]
\begin{center}
\includegraphics[scale=0.25]{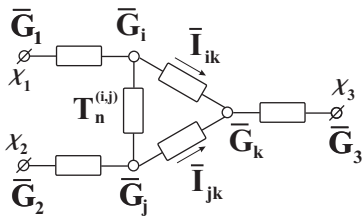}
\caption{The graph of the circuit theory, associated with a 3 terminal
mesoscopic system. 
}
\end{center}
\end{figure}

%METHOD: CIRCUIT THEORY APPROACH
Next, we construct a theory which makes use of this
conservation law. We concentrate on the Green function in coinciding points,
$\bar G(x,\epsilon)\equiv i\bar G(x,x';\epsilon)/\pi\nu $. So defined Green 
function has been introduced in several semiclassical 
theories. \cite{Rammer,Larkin,Noneq} It satisfies the normalization condition 
$\bar G^2=\bar 1$. We relate the "current density" $\bar j$ to gradients and/or
changes of $\bar G(x)$, very much like the electric current density
is related to the voltage in circuit theory of electric conductance.
Following the approach of the circuit theory, we separate a mesoscopic layout
into elements: nodes and connectors, so that the $\bar G(x)$ is constant across 
the nodes and drops across the connectors. One may associate a graph with each 
circuit, so that its lines $(i,j)$ would denote the connectors, and internal 
and external vertices correspond to the nodes and terminals, respectively. 
(See Fig.~1) 
This separation of actual layout is rather heuristic, similar to
separation of an electric conductor of a complicated geometry onto nodes
and circuit theory elements. The bigger the number and the finer the
mesh of the nodes and connectors, the better the circuit theory
approximates the actual layout. The nodes are similar to the terminals,
the difference is that $\bar G$ is fixed in the terminals and yet to be
determined in the nodes. The $\bar G$ in nodes are determined from Kirchoff 
rules reflecting the conservation law (\ref{conservation}): sum of the matrix 
currents from the node over all connectors should equal zero at each energy. 
For this, we should be able to express the matrix current via each connector 
as a function of two  matrices $\bar G_{i,j}$ at its ends.

%METHOD: CONNECTORS
The connector $(i,j)$ can be quite generally 
characterized by a set of transmission eigenvalues 
$T_n^{(ij)}$.\cite{General,Noneq} 
The problem to solve is to express matrix current via the connector in terms of 
$\bar{G}_{i(j)}$. This problem shall be addressed by a more microscopic approach 
and was solved in \cite{Noneq} for Keldysh-Nambu matrix structure of $\check G$. 
It is a good news that the derivation made in \cite{Noneq} does not depend on 
concrete matrix structure and can be used for the present problem without any 
modification yielding
\begin{equation}
\label{current}
  \bar{I}_{ij} =\frac{1}{2\pi}\sum_n\int dE 
  \frac{T^{(ij)}_n\left[\bar G_i,\bar G_j\right]}{
    4+T^{(ij)}_n\left(\{\bar G_i, \bar G_j\} - 2 \right)}\,.
\end{equation}
Each connector $(i,j)$ in the layout contributes to the total $\chi_i$-dependent 
action (\ref{action}). The corresponding $S_{ij}$ contribution 
reads:\cite{Wolfgang}
\begin{equation}
  S_{ij}(\chi)=\frac{-t_0}{2\pi}\sum_n\int dE \text{Tr}
  \ln\left[1+\frac{1}{4} T^{(ij)}_n\left(\{\bar G_i, \bar G_j\}-
2\right)\right]\;.
\label{action-connector}
\end{equation}

%METHOD: GENERAL FORMULA AND HOW TO EVALUATE
Now we are ready to present a set of circuit theory rules that enables
us to evaluate the FCS for an arbitrary mesoscopic layout. 
i. The layout is separated onto terminals, nodes, and connectors.
ii. The $\bar G_{j}$ in each terminal $j$ is fixed by relation (\ref{boundary}) 
thus incorporating information about voltage, temperature and counting field 
$\chi$ in each node. 
iii. For each node $k$, the matrix current conservation yields a Kirchoff equation 
$\sum_i \bar I_{ik}=0$, where the summation is going over all connectors $(i,k)$ 
attached to node $k$, and $\bar I_{ik}$ are expressed with
(\ref{current}) in terms of $\bar G_{i(k)}$.
iv. The solution of resulting equations with condition $\bar G_{k}^2=1$ fixes 
$\bar G_{k}$ in each node.
v. The total action $S(\chi)$ is obtained by summing up the contributions 
$S_{ij}(\{\chi_i\})$ of individual connectors, those are given by 
(\ref{action-connector}):
$
S(\{\chi_i\})=\sum_{(i,j)} S_{ij}(\{\chi_i\})
$
vi. The statistics of electron transfer is obtained from the  
relation (\ref{Probability}).
\label{rules}

%DISCUSSION OF APPLICABILITY AND TIME-DEPENDENT STUFF
It is time to discuss the limits of applicability of the whole scheme. By virtue 
of  semiclassical approach, the mesoscopic fluctuations coming from interference 
of electrons penetrating different connectors are disregarded. So that, we 
assume that conductivities of all connectors are much bigger than conductance 
quantum $e^2/\pi\hbar$. The same condition provides the absence of 
Coulomb blockade effects in the system. Besides, we have disregarded the 
possible processes of {\it inelastic relaxation} in the system. The latter can be 
eventually taken into account but it would 
considerably complicate the scheme. The point is that the inelastic 
scattering would mix up the $\bar G$ at different energies, so that one can not 
solve the circuit theory equations separately at each energy.

\begin{figure}[t]
%\begin{center}
\includegraphics[scale=0.4]{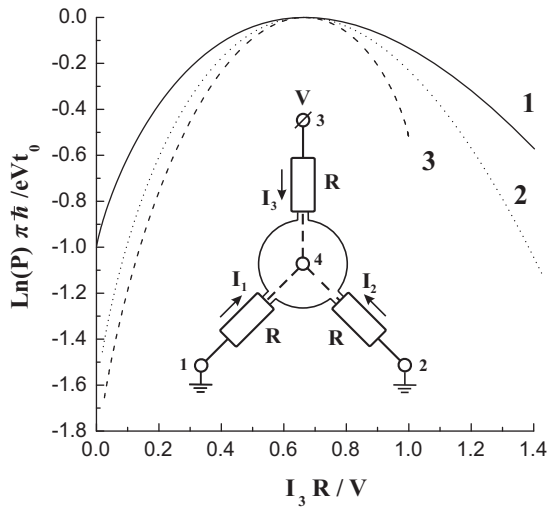}
\caption{The logarithm of the current probabilities in the 3-terminal 
chaotic quantum dot as a function of $I_3$, under condition $I_1=I_2$. 
The insert presents the system configuration. The resistances $R$
of all connectors are assumed to be equal. 1 - tunnel connectors,
2 - diffusive connectors, 3 - ballistic connectors. }
%\end{center}
\end{figure}

%HERE'S DIMA'S PIECE
%%(i)
As an illustration of the presented scheme, we will consider in the rest of the 
paper the FCS of the 3-terminal chaotic quantum dot. The system is sketched 
in the inset of Fig.~2. The heuristic circuit, associated with
this mesoscopic system is shown by dashed lines. It includes only 3 external
vertices, corresponding to terminals, 3 arbitrary connectors, associated with  
the contacts, and the node $\{4\}$, representing the quantum dot itself.
This separation is valid provided the cavity is in the {\it quantum}
chaotic regime. (See~\cite{Aleiner} for definition). This regime corresponds to 
full isotropization of the Green function 
$\check G(x,x',\epsilon)$ within the dot, so that $\bar G_4(\epsilon)$  
can be regarded as a constant at a given energy.

%%(ii)
Since the normalization $\bar G_k^2=1$ holds for each vertex, we
use the parametrization $\bar G_k={\bf g}_k \cdot \bm{\tau}$,
${\bf g}_k \cdot {\bf g}_k =1$.  Here ${\bf g}_k$ is a 3-D vector, and
$\bm{\tau} = (\bar\tau_1,\bar\tau_2,\bar\tau_3)$. In the absence of counting fields
the Green functions in the terminals are given by a zero condition
%\begin{equation}
$
\bar G_k^{(0)} = \left(
\begin{array}{cc}
 1-2f_k & -2 f_k \\
 -2(1-f_k) & 2f_k-1
\end{array} \right),
%\end{equation}
$
where Fermi distribution function $f_k(E)=\{ \exp[(E-eV_k)/T_k] + 1\}^{-1}$
accounts for the bias voltages $V_k$ and the temperatures $T_k$ in the 
terminals.
The $\chi_i$-dependence of $\bar G_k(\chi)$ is then given by Eq.~\ref{boundary}. 

%%{iii)
We see that $\bar G_4(\chi)={\bf g}_4 \cdot \bm{\tau}$ 
is in fact the only function to be found. 
For that, we proceed by applying the current conservation law,
$\sum_{k=1}^3 \bar I_{k,4}=0$, inside the dot. We present 
the currents $\bar I_{k,4}$ given by~(\ref{current}) in the form
%\begin{equation}
$\bar I_{k,4}=\frac{1}{2}Z_k({\bf g}_k \cdot {\bf g}_4)
[\bar G_k,\bar G_4]$,
%\label{dotcurrents}
%\end{equation}
the scalar function $Z_k(x)$ incorporating the information about transmission eigenvalues
in each connector $k$:
$
Z_k(x)\equiv \sum\limits_n T_n^{(k,4)}/[2+T_n^{(k,4)}(x-1)].
$
It can be evaluated for any particular distribution 
$\rho(T)$ of transmission eigenvalues in the given connector and 
completely  defines its scattering properties.
For a example, if we denote $R_0=\pi\hbar/e^2$, then 
$R^{-1}_k=2 R_0^{-1} Z_k(1)$ is an inverse 
resistance of the connector. One can also express the Fano factor
$F=\langle T(1-T) \rangle/\langle T\rangle$ as 
$F=1-2 (d/dx) {\log Z(x)\bigr |}_{x=1}$.
With the use of $Z_k(x)$ the conservation law 
can be efficiently rewritten as 
$[\sum\limits_{k=1}^{3} p^k \bar G_k, \bar G_4]=0$,
where $p^k=Z_k\left({\bf g}_k \cdot {\bf g}_4 \right)$. The latter
enables one to look for the vector ${\bf g}_4$ in the form 
$
{\bf g}_4 = M^{-1}\sum\limits_{k=1}^3\,p^k\, {\bf g}_k,  
$
with $M(\chi)$ being an unknown normalization constant. Using
the condition ${\bf g}_4 \cdot {\bf g}_4=1$ we obtain the set of 
equations 
%%(iv)
\begin{equation}
\label{mapping}
p^i  =  Z_i\Bigl(M^{-1}\sum\limits_{j=1}^3 g_{ij}(\chi)\,p^j\Bigr), \quad
M^2  =  \sum\limits_{i,j=1}^3 g_{ij}(\chi) p^i p^j
\end{equation}
where 
\begin{eqnarray}
g_{ij}(\chi) &=& {\bf g}_i(\chi) \cdot {\bf g}_j(\chi) 
= (1-2f_i)(1-2f_j) \nonumber \\
&+& 2\,e^{i(\chi_i-\chi_j)} f_i(1-f_j) + 2\,e^{-i(\chi_i-\chi_j)} f_j(1-f_i) 
\nonumber
\end{eqnarray}
The fixed point $\{p_{*}^i(\chi), M_*(\chi)\}$ of the mapping~(\ref{mapping}) 
finally 
yields $\bar G_4$ in question.

%%(v)
The total action can be found by applying the rule (v) at page~\pageref{rules}
and reads
\begin{equation}
%$$
S(\chi) = \frac{\displaystyle t_0}{\displaystyle \pi} 
\sum\limits_{i=1}^3 \int d\epsilon\,
S_i\Bigl(M_*^{-1}(\chi)\sum\limits_{j=1}^3 g_{ij}(\chi)\,p_{*}^i(\chi) \Bigr)
\label{FCS-Action}
\end{equation}
%$$
The partial contributions $S_i(x)$ in the above equation should be determined from
the relation $\frac{\partial}{\partial x}S_i(x)=-Z_i(x)$, $S_i(1)=0$.

%% Connectors
We specifically consider three particular types of connectors: 
tunnel(T), ballistic(B) and diffusive(D). 
The corresponding contributions to action are: 
%\begin{eqnarray}
$S_T(x) = -\frac{1}{2}(R_0/R) (x-1),$
$S_B(x) = - (R_0/R) \log[(1+x)/2]$,
$S_D(x) = -\frac{1}{4} (R_0/R) \log^2(x+\sqrt{x^2-1})$~\cite{RefYuli},
$R$ being a resistance of the connector.
For tunnel connector $T_n\ll 1$ for all $n$.
For ballistic connector $N$ channels are opened
($T_n\approx 1$ for $n\le N$), and the other are closed. 
In the diffusive connector the transmission eigenvalues are distributed
according to universal law $\rho(T)= R_0/2RT\sqrt{1-T}$.

%%Numerical scheme.
Analytical results for FCS~(\ref{FCS-Action}) are plausible only for the system 
with tunnel connectors. To assess general situation we solve the
Eqs.~(\ref{mapping}) for given $\chi_i$ numerically.
To find the probability distribution, we evaluated the
integral~(\ref{Probability}) in the saddle point approximation, assuming
$\chi_i$ to be complex numbers. Saddle point approximation is always valid
in the low frequency limit we consider, since in this case both action
$S$ and number of transmitted particles $N_i=I_i t_0/e \gg 1$. 
Due to the current conservation law 
only two of three counting fields ${\chi_i}$ are independent, and one can set 
$\chi_3=0$. The relevant saddle point of the function 
$\Omega(\chi) = S(\chi)+ i\chi_1 I_1 t_0/e + i\chi_2 I_2 t_0/e$ always 
corresponds to purely imaginary numbers $\{\chi_1^*,\chi_2^*\}$ in the upper half plane.
The probability reads $P(I_1,I_2)\approx \exp[-\Omega(\chi^*)]$. Evidently, 
$\Omega(\chi^*)$
is the Legendre transform of the action, and it can be regarded
as implicit function on $I(\chi^*)$.

\begin{figure*}
\includegraphics[scale=0.75]{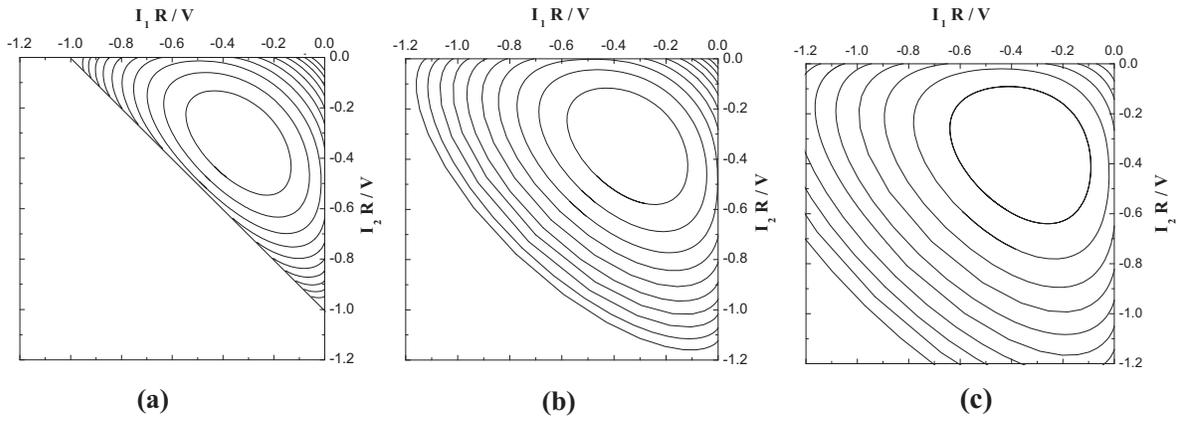}
\caption{The contour maps of the current distribution $\log[P(I_1,I_2)]$ in
the 3-terminal chaotic quantum dot for different configurations of connectors.
(a) - ballistic connectors, (b) - diffusive connectors, (c) - tunnel connectors.}
\end{figure*}

%Results
In the following we assume the short noise regime $eV\gg kT$
when the thermal fluctuations can be disregarded. The energy
integration in~(\ref{FCS-Action}) becomes trivial, since $f_i(\epsilon)=0$ or 1,
and it is sufficient to consider only the case $V_1=V_2=0$, $V_3=V$. 
Any other possible setup can be reduced to the number of previous ones by appropriate 
subdividing a relevant energy strip. The results of these calculations are 
shown in Fig.2 and 3. We see that the maximum of probability  
occurs at $I_1=I_2=-V/3R$, $I_3=2V/3R$ that simply reflects the usual Kirchoff rules. 
The current distribution $P(I_1,I_2)$
for a ballistic system is bounded. It is due to the fact, that $Z_B(x)$ 
contains the finite number of open channels, contrary to the tunnel or diffusive 
type configurations, where it is not the case.  From~(\ref{mapping}) and 
(\ref{FCS-Action}) we can also find a zero noise and noise correlations matrix 
$\tilde S_{ij}= eR^{-1}V F_{ij}$, $F_{11}=F_{22}=(4+3F)/27$, $F_{33}=(4+6F)/27$, 
$F_{12}=-2/27$, $F_{13}=F_{23}=-(2+3F)/27$, where $F$ is a Fano factor.
Since $F_B=0$, $F_D=1/3$ and $F_T=1$ one concludes,
that for a fixed average currents through  connectors the Gaussian's currents 
fluctuations will increase in the sequence
ballistic$\rightarrow$diffusive$\rightarrow$tunnel. Fig.2 and 3 show, that the 
similar behavior is also traced in the regime of the large  current 
fluctuations. The essential point here is that the cross-correlations 
always persist regardless the concrete construction of the connectors. For the case of
multilead chaotic cavities the results  for shot noise in our theory coincides with those, 
obtained by means of  random matrix theory~\cite{Langen}, and with use of 
"minimal correlation approach"~\cite{BlanterSukhor}.

%CONCLUSIONS
In conclusion, we present a constructive theory for counting statistics for 
electron transfer in mesoscopic systems. With this theory, one can easily make 
theoretical predictions for all FCS, thereby facilitating experimental 
activities in this direction. Up to now, only the noise has been measured.
In our opinion, the measurements of FCS can be easily 
performed with {\it threshold detectors} that
produce a signal provided the current in a certain terminal exceeds the 
threshold value. If the threshold value exceeds the average current, the 
detector will be switched by relatively improbable fluctuations of the current 
described by FCS, its signal being proportional to the probability $P(I_1,I_2)$ 
of these fluctuations.

%THANKS
We appreciate useful discussions with G. E. W. Bauer, C. W. J. Beenakker, L. P. 
Kouwenhoven and L. S. Levitov.
This work is a part of the research program of the "Stichting voor
Fundamenteel Onderzoek der Materie"~(FOM), and I acknowledge the financial
support from the "Nederlandse Organisatie voor Wetenschappelijk Onderzoek"~(NWO).

\end{document}